\documentclass[proceedings]{JHEP3}
\usepackage{amsfonts}
\usepackage{amssymb}
\usepackage{amsmath}
\usepackage{epsfig,multicol}

\newbox\mybox

\newcommand\fverb{\setbox\mybox=\hbox\bgroup\verb}
\newcommand\fverbdo{\egroup\medskip\noindent\fbox{\unhbox\mybox}\ }
\newcommand\fverbit{\egroup\item[\fbox{\unhbox\mybox}]}
\conference{Non-Hermitian Hamiltonians of Lie algebraic type}
\abstract{We analyse a class of non-Hermitian Hamiltonians, which
can be expressed in terms of bilinear combinations of generators in the $sl_{2}(\mathbb{R})$-Lie algebra or their isomorphic $su(1,1)$-counterparts. The Hamlitonians are prototypes for solvable models of Lie algebraic
type. Demanding a real spectrum and the existence of a well defined metric,
we systematically investigate the constraints these requirements 
impose on the coupling constants of the model and the parameters in the metric
operator. We compute isospectral Hermitian counterparts for some of the
original non-Hermitian Hamiltonian. Alternatively we employ a generalized Bogoliubov transformation, which allows to compute 
explicitly real energy eigenvalue spectra for these type of Hamiltonians, 
together with their eigenstates. We compare the two approaches.}

\title{Non-Hermitian Hamiltonians of Lie algebraic type}
\author{Paulo E.G. Assis and Andreas Fring \\
Centre for Mathematical Science, City University London, \\
Northampton Square, London EC1V 0HB, UK\\
E-mail: Paulo.Goncalves-De-Assis.1@city.ac.uk, A.Fring@city.ac.uk}

\input{tcilatex}

\begin{document}

\section{Introduction}

Non-Hermitian operators in a complex Hilbert space have been studied in the
mathematical literature \cite{Heu,Dieu,Sheth,Will} for a long time. Also in
various contexts of physics non-Hermitian Hamiltonians have frequently
occurred over the years. Besides those having complex eigenvalue spectra,
and thus describing dissipative systems, some with real eigenvalues have
been considered too. For instance in the study of strong interactions at
high energies in form of Regge models \cite{Regge,Regge2}, in integrable
quantum field theories in form of affine Toda field theories with complex
coupling constants \cite{Holl,David}, in condensed matter physics in the
context of the XXZ-spin chain \cite{Korff:2007qg} and recently also in a
field theoretical scenario in the quantization procedure of strings on an $%
AdS_{5}\times S^{5}$- background \cite{Das:2007tb}. Various attempts to
understand these sort of Hamiltonians have been made over the years, e.g.~%
\cite{Dieu,Sheth,Will,Cal,Urubu}, which may be traced back more than half a
century \cite{Heisenberg}. A more systematic study and revival of such type
of Hamiltonians was initiated roughly ten years ago \cite{Bender:1998ke},
for reviews and special issues see e.g.~\cite%
{special,specialCzech,CArev,Benderrev,special2}. Meanwhile some concrete
experimental settings have been proposed in which properties of these models
can be tested \cite{Exp}. Here we wish to focus on the large subclass of
non-Hermitian Hamiltonians with real eigenvalues of Lie algebraic type.

Many interesting and important physical Hamiltonians may be cast into a Lie
algebraic formulation. For very general treatments one can take these
formulations as a starting point and generic frameworks, such that
particular models simply result as specific choices of representations%
\footnote{%
See section 5 for concrete examples, such as the BCS-Hamiltonian of
supersymmetry and others.}. The virtue of this kind of approach is that it
allows for a high degree of universality and it has turned out to be
especially fruitful in the context of integrable and solvable models, the
former implying that the amount of conserved quantities equals the degrees
of freedom in the system and the latter referring to a situation in which
the spectra can be determined explicitly. Here we will extend such type of
treatment to pseudo-Hermitian or more precisely, and more useful, to
quasi-Hermitian Hamiltonian systems. More specifically, we wish to consider
non-Hermitian Hamiltonian systems for which an exact similarity
transformation can be explicitly constructed, such that it transforms it
into a Hermitian one. We refer to them as \emph{solvable quasi Hermitian}
(SQH) Hamiltonian systems, see e.g.~\cite%
{Mist,JM,CA,KBZ,Mostdel,MGH,PEGA,Bender:2008uu} for explicit models of this
type. The main virtues of these models is that they obviously possess real
eigenvalue spectra \cite{Will}, due to the fact that quasi-Hermitian systems
are directly related to Hermitian Hamiltonian systems. Alternatively, and
very often equivalently, one may explain the reality of the spectra of some
non-Hermitian Hamiltonians when one encounters unbroken $\mathcal{PT}$%
-symmetry, which in the recent context was first pointed out in \cite%
{Bender:2002vv}. Unbroken specifies here that both the Hamiltonian and the
wavefunction remain invariant under a simultaneous parity transformation $%
\mathcal{P}:x\rightarrow -x$ and time reversal $\mathcal{T}:t\rightarrow -t$%
. Noting that the $\mathcal{PT}$-operator is just a specific example of an
anti-linear operator this is known for a long time \cite{EW}.

Our manuscript is organised as follows: In section 2 we introduce the basic
ideas of Hamiltonians of Lie algebraic type, focussing especially on the two
isomorphic cases $sl_{2}(\mathbb{R})$ and $su(1,1)$. Section 3 is devoted to
the systematic construction of similarity transformations towards
isospectral Hermitian counterparts and metric operators. In section 4 we
employ generalized Bogoliubov transformations to compute real eigenvalue
spectra for the Hamiltonians of Lie algebraic type and compare our results
with the findings of section 3. In section 5 we comment on some explicit
realisations, which are useful in order to relate to some specific physical
models. Our conclusions are stated in section 6.

\section{Hamiltonians of Lie algebraic type}

The notion of \emph{quasi-exactly solvable}\textit{\ }operators was
introduced by Turbiner \cite{Tur0} demanding that their action on the space
of polynomials leaves it invariant. More specifically when taking the
operator to be a Hamiltonian operator $H$ acting on the space of polynomials
of order $n$ as $H$: $V_{n}\mapsto V_{n}$, it preserves by definition the
entire flag $V_{0}\subset V_{1}\subset V_{2}\subset \ldots \subset
V_{n}\subset \ldots ~$Models respecting this property are referred to as 
\emph{exactly solvable}. Whenever these type of Hamiltonians can be written
in terms of bilinear combinations of first-order differential operators
generating a finite dimensional Lie algebra, it is said they are of \emph{%
Lie algebraic type} \cite{Olv}.

In order to be more concrete we have to identify $V_{n}$ as the
representation space of some specific Lie algebra. The simplest choice is to
involve the only rank one Lie algebra $sl_{2}(\mathbb{C})$. It is well known
that this algebra contains the compact real form $su(2)$ and the non-compact
real form $sl_{2}(\mathbb{R})$, which is isomorphic to $su(1,1)$, see for
instance \cite{Hum,repr}. We will focus here on these two choices.

\subsection{Hamiltonians of $sl_{2}(\mathbb{R})$-Lie algebraic type}

The three generator $J_{0}$, $J_{1}$ and $J_{2}$ of $sl_{2}(\mathbb{R})$
satisfy the commutation relations $\left[ J_{1},J_{2}\right] =-iJ_{0}$, $%
\left[ J_{0},J_{1}\right] =iJ_{2}$ and $\left[ J_{0},J_{2}\right] =-iJ_{1}$,
such that the operators $J_{0}$, $J_{\pm }=J_{1}\pm J_{2}$ obey 
\begin{equation}
\left[ J_{0},J_{\pm }\right] =\pm J_{\pm },\qquad \left[ J_{+},J_{-}\right]
=-2J_{0},\qquad \text{and\qquad }J_{0}^{\dagger },J_{\pm }^{\dagger }\notin
\{J_{0},J_{\pm }\}.  \label{ss}
\end{equation}%
As possible realisation for this algebra one may take for instance the
differential operators 
\begin{equation}
J_{-}=\partial _{x},\text{\qquad }J_{0}=x\partial _{x}-\frac{n}{2},\qquad
J_{+}=x^{2}\partial _{x}-nx,~~~~n\in \mathbb{Z}\text{,}  \label{JJJ}
\end{equation}%
allegedly attributed to Sophus Lie, see e.g.~\cite{Tur0}. Clearly the action
of this algebra on the space of polynomials 
\begin{equation}
V_{n}=\limfunc{span}\{1,x,x^{2},x^{3},x^{4},...,x^{n}\}
\end{equation}%
leaves it invariant. According to the above specified notions, a\textit{\ }%
quasi-exactly solvable Hamiltonian of Lie algebraic type is therefore of the
general form 
\begin{equation}
H_{J}=\dsum\limits_{l=0,\pm }\kappa _{l}J_{l}+\dsum\limits_{n,m=0,\pm
}\kappa _{nm}:J_{n}J_{m}:,\qquad \kappa _{l},\kappa _{nm}\in \mathbb{R}\text{%
,}  \label{HJ}
\end{equation}%
where we introduced the ordering 
\begin{equation}
:J_{n}J_{m}:=\left\{ 
\begin{array}{r}
J_{n}J_{m}\qquad \text{for }n\geq m \\ 
0\qquad \text{for }n<m%
\end{array}%
\right.  \label{ord}
\end{equation}%
to avoid unnecessary double counting\footnote{%
By setting some of the arrangements to zero our normal ordering prescription
differs slightly from the ordinary one, but this is simply convention here
and has no bearing on our analysis.}. This means the Hamiltonian $H_{J}$
involves nine real constants $\kappa $, plus a possible overall shift in the
energy. It is evident from the representation (\ref{JJJ}) that when $\kappa
_{+}=\kappa _{++}=\kappa _{+0}=0$ the model becomes exactly solvable in the
sense specified above. For the given representation (\ref{JJJ}) the $%
\mathcal{PT}$-symmetry may be implemented trivially by rescaling $J_{\pm
}\rightarrow \tilde{J}_{\pm }=\pm iJ_{\pm }$ and $J_{0}\rightarrow \tilde{J}%
_{0}=J_{0}$, which leaves the algebra (\ref{ss}) unchanged. Taking the
algebra in this representation will leave the real vector space of $\mathcal{%
PT}$-symmetric polynomials 
\begin{equation}
V_{n}^{\mathcal{PT}}=\limfunc{span}\{1,ix,x^{2},ix^{3},x^{4},...,e^{i\pi
n/2}x^{n}\}
\end{equation}%
invariant. Since by construction the Hamiltonian $H_{\tilde{J}}$ and the
wavefunctions are $\mathcal{PT}$-symmetric, as they are polynomials in $%
V_{n}^{\mathcal{PT}}$, the eigenvalues for these systems must be real by
construction \cite{Bender:2002vv,EW}. Nonetheless, to determine the explicit
similarity transformation remains a challenge.

A simple explicit example for $H_{\tilde{J}}$ with $\kappa _{00}=-4$, $%
\kappa _{+}=-2\zeta =\kappa _{-}$, $\zeta \in \mathbb{R}$ an overall energy
shift by $M^{2}+\zeta ^{2}$ and all remaining coefficients equal to zero was
recently studied by Bagchi et al \cite{Bag1,Bag2}. The Hamiltonian arises as
a gauged version from the $\mathcal{PT}$-symmetric potential $V(x)=-\left[
\zeta \sinh 2x-iM\right] ^{2}$. The first energy levels together with their
corresponding wavefunctions were constructed and the typical real energy
spectrum for unbroken\ $\mathcal{PT}$-symmetry and complex conjugate pairs
for broken $\mathcal{PT}$-symmetry was found. However, even for this simple
version of (\ref{HJ}) a general treatment leading to the complete eigenvalue
spectrum and a well defined metric has not been carried out.

As we indicated, the representation (\ref{JJJ}) is ideally suited with
regard to the question of solvability. However, the Hermiticity properties
for the $J$'s are not straightforward to determine within a Lie algebraic
framework, since the Hermitian conjugates of the $J$'s can not be written in
terms of the original generators. This feature makes the representation (\ref%
{JJJ}) rather unsuitable for the determination of the Hermiticity properties
of the Hamiltonian $H_{\tilde{J}}$ in generality. The implication is that we
may carry out our programme only for specific representations using directly
some concrete operator expressions or equivalently Moyal products of
functions \cite{Moyal2,ACIso,PEGA}, see section 3.1 for an example, and not
in a generic representation independent way. An additional undesired feature
is that the Hamiltonian $H_{\tilde{J}}$ in terms the representation (\ref%
{JJJ}) does not allow to capture many of the important and interesting
physical models. We will therefore consider a slightly different type of
algebra.

\subsection{Hamiltonians of $su(1,1)$-Lie algebraic type}

The above mentioned problems do not occur when we express our Hamiltonian in
terms of the isomorphic $su(1,1)$-Lie algebra, whose generators $K_{0}$, $%
K_{1}$ and $K_{2}$ satisfy the same commutation relations $\left[ K_{1},K_{2}%
\right] =-iK_{0}$, $\left[ K_{0},K_{1}\right] =iK_{2}$ and $\left[
K_{0},K_{2}\right] =-iK_{1}$. Consequently the operators $K_{0}$, $K_{\pm
}=K_{1}\pm K_{2}$ satisfy an isomorphic algebra to (\ref{ss}) 
\begin{equation}
\left[ K_{0},K_{\pm }\right] =\pm K_{\pm },\qquad \left[ K_{+},K_{-}\right]
=-2K_{0}\qquad \text{and\qquad }K_{0}^{\dagger }=K_{0},K_{\pm }^{\dagger
}=K_{\mp }.  \label{comK}
\end{equation}

In analogy to (\ref{HJ}) we may then consider a Hamiltonian of Lie algebraic
type in terms of the $su(1,1)$-generators 
\begin{equation}
H_{K}=\dsum\limits_{l=0,\pm }\mu _{l}K_{l}+\dsum\limits_{n,m=0,\pm }\mu
_{nm}:K_{n}K_{m}:,\qquad \mu _{l},\mu _{nm}\in \mathbb{R},  \label{HK}
\end{equation}%
where we have used the same conventions for the ordering as in equation (\ref%
{ord}). In general this Hamiltonian is not Hermitian, that is when the
constants $\mu _{+}\neq \mu _{-}$, $\mu _{++}\neq \mu _{--}$ or $\mu
_{+0}\neq \mu _{0-}$ we have $H_{K}^{\dagger }\neq H_{K}$. Our main aim is
now to identify a subset of Hamiltonians $H_{K}$, which despite being
non-Hermitian possess a real eigenvalue spectrum.

There are various types of representations in terms of differential
operators for this algebra as for instance the multi-boson representation 
\begin{equation}
K_{0}=k_{0}(N),\qquad \qquad K_{+}=k_{+}(N)(a^{\dag })^{n},\qquad \qquad
K_{-}=k_{-}(N)(a)^{n},  \label{multi}
\end{equation}%
where the $a,a^{\dagger }$ are the usual bosonic annihilation and creation
operators with $N=a^{\dag }a$ being the number operator. The $k_{0}(N)$, $%
k_{\pm }(N)$ are functions of the latter and may be determined recursively
for any number of bosons $n$ involved \cite{goli}. The simplest case $n=1$
yields the Holstein-Primakoff representation \cite{Holst} with $K_{0}=N+%
\frac{1}{2}$, $K_{+}=\sqrt{N}a^{\dag }$ and $K_{-}=a\sqrt{N}$. For $n=2$ one
obtains the very well known two boson representation 
\begin{equation}
K_{0}=\frac{1}{2}(a^{\dag }a+\frac{1}{2}),\qquad \qquad K_{+}=\frac{1}{2}%
a^{\dag }a^{\dag },\qquad \qquad K_{-}=\frac{1}{2}aa.  \label{bosrep}
\end{equation}

Differential operators in $x$-space are then obtained by the usual
identification $a=(\omega \hat{x}+i\hat{p})/\sqrt{2\omega }$ and $a^{\dagger
}=(\omega \hat{x}-i\hat{p})/\sqrt{2\omega }$ with the operators $\hat{x}$, $%
\hat{p}=-i\partial _{x}$ and $\omega \in \mathbb{R}.$

The part of the Hamiltonian $H_{K}$ linear in the generators $K$ corresponds
to the Hamiltonian recently studied by Quesne \cite{Quesne}, who constructed
an explicit metric operator for this Hamiltonian together with its Hermitian
isospectral partner. For the particular representation (\ref{bosrep}) this
reduces to the so-called Swanson Hamiltonian \cite{Swanson}, for which
various metric operators were constructed previously by Musumbu et al \cite%
{MGH}. Here we shall extend the analysis to the case involving bilinear
combinations, staying as generic as possible without appealing to any
particular representation.

\section{Construction of a metric operator and Hermitian counterpart\label%
{met}}

Our key aim is now to construct a well defined metric operator, i.e. a
linear, invertible, Hermitian and positive operator acting in the Hilbert
space, such that $H$ becomes a self-adjoint operator with regard to this
metric. Our starting point will be the assumption that there exists a
similarity transformation, which maps the non-Hermitian Hamiltonian $H$
adjointly to a Hermitian Hamiltonian $h$ 
\begin{equation}
h=\eta H\eta ^{-1}=h^{\dagger }=(\eta ^{-1})^{\dagger }H^{\dagger }\eta
^{\dagger }~~\Leftrightarrow ~~H^{\dagger }\rho =\rho H\text{ \ with }\rho
=\eta ^{\dagger }\eta \text{.}  \label{1}
\end{equation}%
There exist also variations of these properties, which lead to less
stringent conclusions. We summarize the most common ones in the following
table

\begin{center}
\begin{tabular}{|l|c|c|c|}
\hline
& $H^{\dagger }=\eta ^{\dagger }\eta H(\eta ^{\dagger }\eta )^{-1}$ & $%
H^{\dagger }\rho =\rho H$ & $H^{\dagger }=\rho H\rho ^{-1}$ \\ \hline
positivity of $\rho $ & $\checkmark $ & $\checkmark $ & $\times $ \\ \hline
Hermiticity of $\rho $ & $\checkmark $ & $\checkmark $ & $\checkmark $ \\ 
\hline
invertibility of $\rho $ & $\checkmark $ & $\times $ & $\checkmark $ \\ 
\hline
terminology for $H$ & (\ref{1}) & quasi-Hermiticity\cite{Dieu} & 
pseudo-Hermiticity\cite{pseudo1} \\ \hline
spectrum of $H$ & real\cite{Will} & could be real\cite{Dieu} & real\cite%
{Will} \\ \hline
definite metric & guaranteed & guaranteed & not conclusive \\ \hline
\end{tabular}
\end{center}

We should stress that this is the most frequently used terminology and at
times it is mixed up and people imply different properties by using the same
names. Making no assumption on the positivity of the $\rho $ in (\ref{1}),
the relation on the right hand side constitutes the well known \emph{%
pseudo-Hermiticity} condition, see e.g.~\cite%
{pseudo1,pseudo2,Mostafazadeh:2001nr}, when the operator $\rho $ is linear,
invertible and Hermitian. In case the operator $\rho $ is positive but not
invertible this condition is usually referred to as \emph{quasi-Hermiticity}%
\footnote{%
Surprisingly the early literature on the subject, such as \cite%
{Dieu,Sheth,Will}, is entirely ignored in recent publications and statements
such as "the terminology quasi Hermitian was coined in \cite{Urubu}", see
e.g.~\cite{UrubuZ}, are obviously incorrect. The term \textit{%
quasi-Hermitian operator} was first introduced by Dieudonn\'{e} in 1960 \cite%
{Dieu}. Relaxing the requirement of invertibility the operators become 
\textit{symmetrizable operators} for which there exists an extensive even
earlier literature, see e.g.~\cite{Heu} and references therein.} \cite%
{Dieu,Sheth,Will,Urubu}. With regard to the properties of discrete spectra
of $H$ the difference is irrelevant as both conditions may be used to
establish its reality. However, in the case of pseudo-Hermiticity this is
guaranteed whereas in the case of quasi-Hermiticity one merely knowns that
it could be real. With regard to the construction of a metric operator the
difference becomes also important, since pseudo-Hermiticity may lead to an
indefinite metric, whereas quasi-Hermiticity will guarantee the existence of
a positive definite metric.

Naturally we expect to find many solutions if the Hermitian Hamiltonian $h$
is not specified concretely. In other words when given the non-Hermitian
Hamiltonian $H$ the similarity transformation is not unique when the only
requirement for $h$ is its Hermiticity. The ambiguities indicate the
existence of a symmetry, see e.g.~section 3.3 in \cite{PEGA}. However,
uniqueness may be achieved by specifying either the concrete form of $h$ or
any other irreducible observable \cite{Urubu}.

\subsection{Hamiltonians of $sl_{2}(\mathbb{R})$-Lie algebraic type}

Let us start by considering first the Hamiltonian $H_{\tilde{J}}$ in (\ref%
{HJ}) in the context of the above mentioned programme and try to solve the
equation

\begin{equation}
h_{\tilde{J}}=\eta H_{\tilde{J}}\eta ^{-1}=h_{\tilde{J}}^{\dag }  \label{hhj}
\end{equation}%
for $\eta $. As a general ansatz we start with the non-Hermitian operator

\begin{equation}
\eta =e^{2\varepsilon \lbrack \tilde{J}_{0}+\lambda (\tilde{J}_{+}+\tilde{J}%
_{-})]}\neq \eta ^{\dagger }\qquad \varepsilon ,\lambda \in \mathbb{R}.
\label{etan}
\end{equation}%
Unlike as in the case when $\eta =\eta ^{\dagger }$, see section 3.2, we do
not need to worry here about the positivity of $\eta $, since the
decomposition of the metric operator $\rho =\eta ^{\dagger }\eta $ ensures
it to be positive.

One of the simplest cases to consider for expressions of $H_{\tilde{J}}$ is
the purely linear one, i.e. when all $\kappa _{nm}$ vanish. In principle
this Hamiltonian fits into the class of general $\mathcal{PT}$-symmetric
Hamiltonians considered in \cite{PEGA}, when the constants therein are
identified as $\alpha _{3}=-\kappa _{+}$, $\alpha _{4}=-\kappa _{-}$, $%
\alpha _{6}=-(n+1)\kappa _{0}/2$, $\alpha _{8}=\kappa _{0}$, $\alpha
_{10}=-(n+1)\kappa _{+}$ and all remaining constants are taken to be zero.
However, none of the exactly solvable models obtained in there matches with $%
H_{\tilde{J}}$. Nonetheless, relaxing the condition $\eta =\eta ^{\dagger }$
as in the ansatz (\ref{etan}) allows to construct an exact Hermitian
isospectral counterpart. An example of how to transform the non-Hermitian
Hamiltonian $H_{\tilde{J}}$ to a Hermitian Hamiltonian is given when the
parameters in the model are related as

\begin{equation}
\kappa _{0}=\pm 2\sqrt{\kappa _{-}\kappa _{+}}\qquad \text{and\qquad }\frac{%
\tanh {\chi }}{\chi /\varepsilon }=\pm \frac{\sqrt{\kappa _{+}}}{\sqrt{%
\kappa _{+}}+2\lambda \sqrt{\kappa _{-}}},
\end{equation}%
with $\chi ={\varepsilon \sqrt{1-4\lambda ^{2}}}$. The Hermitian Hamiltonian
counterpart is subsequently computed to

\begin{equation}
h_{\tilde{J}}=\left( \pm \frac{1}{2\lambda }\kappa _{0}+\kappa _{+}+\kappa
_{-}\right) \tilde{J}_{-}.
\end{equation}

Another interesting simple example is obtained by setting all terms
involving the generator $\tilde{J}_{+}$ to zero, that is taking $\kappa
_{+}=\kappa _{++}=\kappa _{+0}=\kappa _{+-}=0$. In this case we are led to
the relations

\begin{equation}
\kappa _{0}=-(n+1)\kappa _{00},\;\;\;\;\;\;\kappa _{-}=-\frac{n}{\lambda }%
\mu _{00},\;\;\;\;\;\;\kappa _{--}=\frac{\mu _{00}}{\lambda ^{2}}%
,\;\;\;\;\;\;\kappa _{0-}=\frac{2}{\lambda }\kappa _{00}
\end{equation}%
together with

\begin{equation}
\tanh {\chi }=\chi /\varepsilon .
\end{equation}%
The non-Hermitian Hamiltonian $H_{\tilde{J}}$ is then transformed to the
Hermitian Hamiltonian

\begin{equation}
h=\kappa _{00}\tilde{J}_{0}^{2}-\kappa _{0}\tilde{J}_{0}
\end{equation}%
with $0<|\lambda |<\frac{1}{2}$. These examples demonstrate that it is
possible to carry out the above mentioned programme for some specific
realisations of the $sl_{2}(\mathbb{R})$-Lie algebra, albeit not in complete
generality and in a generic representation independent manner.

\subsection{Hamiltonians of $su(1,1)$-Lie algebraic type}

We shall now see that the Hamiltonians $H_{K}$ in (\ref{HK}) allow for a
more general treatment as the problems of the previous section may be
circumvented. In analogy to (\ref{hhj}) let us therefore solve the equation 
\begin{equation}
h_{K}=\eta H_{K}\eta ^{-1}=h_{K}^{\dagger }  \label{herm}
\end{equation}%
for $\eta $.

To start with we take a similar operator ansatz for the similarity
transformation as the one chosen in \cite{MGH,Quesne} 
\begin{equation}
\eta =\exp (2\varepsilon K_{0}+2\nu _{+}K_{+}+2\nu _{-}K_{-}),  \label{eta}
\end{equation}%
where the parameters $\varepsilon ,\nu _{+},\nu _{-}$ are left variable for
the time being. Hermiticity for this operator $\eta $ may be guaranteed when
we take from the very beginning $\nu _{+}=\nu $, $\nu _{-}=\nu ^{\ast }$ and 
$\varepsilon \in \mathbb{R}$ together with the Hermiticity conditions for
the Lie algebraic generators as specified in (\ref{comK}). Noting that the
eigenvalue spectrum of $\eta $ is given by $\exp [(n+1/2)\sqrt{\varepsilon
^{2}-4\nu _{+}\nu _{-}}]$ one can ensure the positivity of $\eta $ when $%
\varepsilon ^{2}>4\nu _{+}\nu _{-}$. One should note that these properties
are only essential for the metric operator $\eta ^{\dagger }\eta $. making
now the more restrictive assumption that $\eta $ is Hermitian, we can obtain
a linear, invertible positive Hermitian metric operator $\eta ^{2}$ when $%
\varepsilon ^{2}-4\left\vert \nu \right\vert ^{2}>0$. Following \cite{MGH}
it is convenient to introduce the variable $\theta =\sqrt{\varepsilon
^{2}-4|\nu |^{2}}$.

Besides the restriction we impose on $\eta $ by demanding it to be
Hermitian, we could have been also more generic by making a more general
ansatz for the expressions for $\eta $, such as for instance allowing in
addition bilinear combinations in the arguments of the exponential. In fact,
as we will show in section \ref{Bogo}, we are certain that more general
types of metric operators must exist. Another very natural version of this
ansatz would be to start with a Gauss or Iwasawa decomposed expression for $%
\eta $.

Using the ansatz (\ref{eta}), we have to compute its adjoint action on $%
H_{K} $ in order to solve (\ref{herm}). In fact, the adjoint action of $\eta 
$ on each of the $su(1,1)$-generators can be computed exactly. We find 
\begin{equation}
\eta K_{l}\eta ^{-1}=t_{l0}K_{0}+t_{l-}K_{-}+t_{l+}K_{+}\qquad \text{for }%
l=0,\pm ,  \label{etaK}
\end{equation}%
where the constant coefficients are 
\begin{eqnarray}
t_{00} &=&1-8|\nu |^{2}\frac{{\sinh {\theta }}^{2}}{{\theta }^{2}},\quad
t_{\pm \pm }=\left( \cosh {\theta }\pm {\varepsilon }\frac{\sinh {\theta }}{%
\theta }\right) ^{2},\quad t_{\pm \mp }=4(\nu _{\pm })^{2}\frac{{\sinh {%
\theta }}^{2}}{{\theta }^{2}}, \\
t_{0\pm } &=&\mp 2\nu _{\mp }\frac{\sinh {\theta }}{\theta }\left( \cosh {%
\theta }\pm {\varepsilon }\frac{\sinh {\theta }}{\theta }\right) ,\qquad
t_{\pm 0}=\pm 4\nu _{\pm }\frac{\sinh {\theta }}{\theta }\left( \cosh {%
\theta }\pm {\varepsilon }\frac{\sinh {\theta }}{\theta }\right) .  \notag
\end{eqnarray}%
These expressions agree with the result in \cite{Quesne}.

With the help of these exact relations we evaluate the adjoint action of $%
\eta $ on the Hamiltonian $H_{K}$ 
\begin{equation}
\eta H_{K}\eta ^{-1}=\dsum\limits_{l=0,\pm }\hat{\mu}_{l}K_{l}+\dsum%
\limits_{n,m=0,\pm }\hat{\mu}_{nm}:K_{n}K_{m}:.  \label{hham}
\end{equation}%
It is evident from (\ref{etaK}) that the general structure of the
Hamiltonian will not change, albeit with a different set of constants $\hat{%
\mu}$, which are rather lengthy and we will therefore not report them here
explicitly. However, they simplify when we impose the constraint that the
resulting Hamiltonian ought to be Hermitian. The condition (\ref{herm})
leads to the six constraints 
\begin{eqnarray}
\hat{\mu}_{0} &=&\hat{\mu}_{0}^{\ast },\qquad \hat{\mu}_{00}=\hat{\mu}%
_{00}^{\ast },\qquad \hat{\mu}_{+-}=\hat{\mu}_{+-}^{\ast },  \label{con1} \\
\hat{\mu}_{+} &=&\hat{\mu}_{-}^{\ast },\qquad \hat{\mu}_{++}=\hat{\mu}%
_{--}^{\ast },\qquad \hat{\mu}_{+0}=\hat{\mu}_{0-}^{\ast }.  \label{con2}
\end{eqnarray}%
The first set of three equations (\ref{con1}) on the reality of $\hat{\mu}%
_{0}$, $\hat{\mu}_{00}$ and $\hat{\mu}_{+-}$ is simply satisfied by the
condition $\nu =\nu ^{\ast }$. Introducing the variables 
\begin{equation}
\lambda =\frac{\nu }{\varepsilon }\qquad \text{and\qquad }Y=\varepsilon 
\frac{\tanh \theta }{\theta }
\end{equation}%
the remaining three equations (\ref{con2}) may be converted into simpler,
albeit still lengthy, equations 
\begin{eqnarray}
0 &=&\mu _{+}-\mu _{-}+2\,Y\,\left[ \mu _{+}+\mu _{-}+2\,\lambda \,\left(
\mu _{++}+\mu _{--}-\mu _{0}-\mu _{00}\right) \right]  \label{cc1} \\
&&+12\,Y^{2}\,\lambda \,\left[ \mu _{++}-\mu _{--}+\lambda \,\left( \mu
_{0-}-\mu _{+0}\right) \right]  \notag \\
&&-2\,Y^{3}\,\left\{ \mu _{+}+\mu _{-}-2\,\lambda \,\left[ \mu _{0}+\mu
_{00}+3\,\left( \mu _{--}+\mu _{++}\right) \right] \right.  \notag \\
&&-{\lambda }^{2}\,\left[ 8\,\mu _{0-}-4\,\left( \mu _{-}+\mu _{+}-2\,\mu
_{+0}\right) \right] +\left. 8\,{\lambda }^{3}\,\left( \mu _{++}+\mu
_{--}+\mu _{0}-\mu _{00}-2\,\mu _{+-}\right) \right\}  \notag \\
&&+Y^{4}\,\left( 1-4\,{\lambda }^{2}\right) \,\left\{ \mu _{-}-\mu
_{+}+4\,\lambda \,\left[ \mu _{++}-\mu _{--}+\lambda \,\left( \mu _{0-}-\mu
_{-}+\mu _{+}-\mu _{+0}\right) \right] \right\} ,  \notag
\end{eqnarray}%
\begin{eqnarray}
0 &=&\mu _{++}-\mu _{--}-2\,Y\,\left[ \lambda \,\left( \mu _{0-}+\mu
_{+0}\right) -2\,\left( \mu _{--}+\mu _{++}\right) \right]  \label{cc2} \\
&&+6\,Y^{2}\,\left[ \mu _{++}-\mu _{--}+\lambda \,\left( \mu _{0-}-\mu
_{+0}\right) \right]  \notag \\
&&-2\,Y^{3}\,\left[ 3\,\lambda \,\left( \mu _{+0}+\mu _{0-}\right) +4\,{%
\lambda }^{3}\,\left( \mu _{+0}+\mu _{0-}\right) -8\,{\lambda }^{2}\,\left(
\mu _{00}+\mu _{+-}\right) -2\,\left( \mu _{++}+\mu _{--}\right) \right] 
\notag \\
&&+Y^{4}\,\left( 1-4\,{\lambda }^{2}\right) \,\left\{ \mu _{++}-\mu
_{--}-2\,\lambda \,\left[ \mu _{+0}-\mu _{0-}+2\,\lambda \,\left( \mu
_{--}-\mu _{++}\right) \right] \right\} ,  \notag
\end{eqnarray}%
\begin{eqnarray}
0 &=&\mu _{+0}-\mu _{0-}+2\,Y\,\left[ \mu _{0-}+\mu _{+0}+4\,\lambda
\,\left( \mu _{++}+\mu _{--}-\mu _{00}-\mu _{+-}\right) \right]  \label{cc3}
\\
&&+24\,Y^{2}\,\,\left[ \lambda (\mu _{++}-\mu _{--})+\lambda ^{2}\,\left(
\mu _{0-}-\mu _{+0}\right) \right]  \notag \\
&&-2\,Y^{3}\,\left\{ \mu _{+0}+\mu _{0-}-4\,\lambda \,\left[ \mu _{00}+\mu
_{+-}+3(\mu _{++}+\mu _{--})\right] -12\lambda ^{2}\,\left( \mu _{0-}+\mu
_{+0}\right) \right.  \notag \\
&&+\left. \,16\lambda ^{3}\,\left( \mu _{++}+\mu _{--}-\mu _{00}-\mu
_{+-}\right) \right\}  \notag \\
&&+Y^{4}\,\left( 1-4\,{\lambda }^{2}\right) \,\left\{ \mu _{0-}-\mu
_{+0}+4\,\lambda \,\left[ \lambda \,\left( \mu _{0-}-\mu _{+0}\right)
+2\,\left( \mu _{++}-\mu _{--}\right) \right] \right\} .  \notag
\end{eqnarray}%
We will now systematically discuss the solutions for these three equations
together with their implications on the metric operator and the
corresponding isospectral pairs of Hamiltonians.

\subsubsection{Non-Hermitian linear term and Hermitian bilinear combinations}

The simplest modification with regard to the purely linear case, treated
previously in \cite{MGH,Quesne}, is to perturb it with Hermitian bilinear
combinations. This means we may assume the equalities $\mu _{++}=\mu _{--}$
and $\mu _{+0}=\mu _{0-}$ in order to determine the relations between the
remaining constants from (\ref{cc1}), (\ref{cc2}) and (\ref{cc3}). We find
that (\ref{cc2}) and (\ref{cc3}) are solved solely by demanding 
\begin{equation}
\mu _{++}=\mu _{--}=\frac{\lambda ^{2}(\mu _{00}+\mu _{+-})}{1+2\lambda ^{2}}%
\;\;\;\;\;\;\;\;\text{and}\;\;\;\;\;\;\;\;\mu _{+0}=\mu _{0-}=\frac{2\lambda
(\mu _{00}+\mu _{+-})}{1+2\lambda ^{2}}\text{,}  \label{313}
\end{equation}%
without any further constraint on $Y$. Solving subsequently equation (\ref%
{cc1}) for $Y$ yields the constraint 
\begin{equation}
\frac{\tanh 2\theta }{\theta /\varepsilon }=\frac{\lambda (\mu _{-}-\mu _{+})%
}{\lambda (\mu _{-}+\mu _{+})+2\lambda ^{2}(\mu _{+-}-\mu _{0})-2\mu _{++}}.
\label{321}
\end{equation}%
Considering (\ref{eta}) we note that the positivity of $\eta ^{2}$ requires $%
\left\vert \lambda \right\vert <1/2$ as a further restriction on the domain
of $\lambda $. Notice that when we send all coefficients $\mu _{nm}$ with $%
n,m\in \{0,\pm \}$ resulting from bilinear combinations to zero we recover
precisely the constraint found in \cite{MGH}, see equation (9) therein.
These equations parametrize the metric and are enough to compute the
Hermitian counterpart via equation (\ref{herm}). We will not report the
expression here as they are rather lengthy and can be obtained as a
reduction from the more general setting to be treated below.

\subsubsection{Hermitian linear term and non-Hermitian bilinear combinations 
\label{HLNHB}}

Reversing the situation of the preceding subsection we may consider the
Hamiltonian $H_{K}$ with Hermitian linear part, i.e. $\mu _{+}=\mu _{-}$,
and non-Hermitian part involving bilinear combinations. In this case we can
solve the equations (\ref{cc1}), (\ref{cc2}) and (\ref{cc3}) by 
\begin{eqnarray}
\mu _{+} &=&\mu _{-}=\lambda (\mu _{0}+\mu _{00}-\mu _{++}-\mu _{--}), \\
\lambda (\mu _{+0}-\mu _{+}) &=&\lambda ^{2}(\mu _{+-}-\mu _{0})+\mu _{++},
\\
\lambda (\mu _{0-}-\mu _{-}) &=&\lambda ^{2}(\mu _{+-}-\mu _{0})+\mu _{--},
\end{eqnarray}%
together with 
\begin{equation}
\frac{\tanh 2\theta }{\theta /\varepsilon }=\frac{\mu _{++}-\mu _{--}}{%
2\lambda \mu _{+}+2\lambda ^{2}(\mu _{+-}-\mu _{0})-(\mu _{++}+\mu _{--})}.
\end{equation}%
This case does not reduce to any case treated in the literature before.

Let us now embark on the general setting in which the linear as well as the
terms in $H_{K}$ involving bilinear combinations are taken to be
non-Hermitian. We will find two different types of solutions, one being
reducible to the foregoing two cases and the other being intrinsically
non-Hermitian and not reducible to any of the previous cases. Reducible is
meant in the sense that the limit of the relevant parameters going to zero
is well defined.

\subsubsection{Generic non-Hermitian reducible Hamiltonian\label{NHLHB}}

Taking now $H_{K}$ to be genuinely non-Hermitian, we find that the equations
(\ref{cc1}), (\ref{cc2}) and (\ref{cc3}) are solved subject to the three
constraints 
\begin{eqnarray}
\mu _{++}-\mu _{--} &=&\lambda (\mu _{+0}-\mu _{0-})  \label{C1} \\
\mu _{--}-\lambda \mu _{0-} &=&\lambda ^{2}(\mu _{++}+\mu _{--}-\mu
_{+-}-\mu _{00})  \notag  \label{C2} \\
2\mu _{+}\mu _{--}-\mu _{-}(\mu _{++}+\mu _{--}) &=&\lambda \lbrack (\mu
_{++}-\mu _{--})(\mu _{++}+\mu _{--}-\mu _{0}-\mu _{00})+\mu _{0-}(\mu
_{+}-\mu _{-})]~  \notag  \label{C3}
\end{eqnarray}%
together with 
\begin{equation}
\frac{\tanh 2\theta }{\theta /\varepsilon }=\frac{\lambda (\mu _{-}-\mu
_{+})+\mu _{++}-\mu _{--}}{\lambda (\mu _{-}+\mu _{+})+2\lambda ^{2}(\mu
_{+-}-\mu _{0})-(\mu _{++}+\mu _{--})}.  \label{T1}
\end{equation}%
We note that taking $\mu _{++}=\mu _{--}$ and $\mu _{+0}=\mu _{0-}$ or $\mu
_{+}=\mu _{-}$ these constraints reduce precisely to the ones previously
treated in the sections \ref{HLNHB} or \ref{NHLHB}, respectively. A further
interesting specialization of this general case is the one involving purely
bilinear combinations, which may be obtained for $\mu _{-}=\mu _{+}=\mu _{0}$
in (\ref{C1}) and (\ref{T1}). For the situation in which the Hamiltonian
does not contain any generators of the type $K_{-}$, i.e. $\mu _{-}=\mu
_{--}=\mu _{0-}=0$ we find 
\begin{equation}
\lambda \mu _{+0}=\mu _{++},\quad \mu _{00}=\mu _{++}-\mu _{0},\quad \mu
_{+-}=\mu _{0},\quad \varepsilon =\frac{\func{arctanh}\sqrt{1-4\lambda ^{2}}%
}{2\sqrt{1-4\lambda ^{2}}},  \label{1minus}
\end{equation}%
and when $H_{K}$ does not contain any generators of the type $K_{+}$, i.e. $%
\mu _{+}=\mu _{++}=\mu _{0+}=0$ the equations simplify to 
\begin{equation}
\lambda \mu _{0-}=\mu _{--},\quad \mu _{00}=\mu _{--}-\mu _{0},\quad \mu
_{+-}=\mu _{0},\quad \varepsilon =-\frac{\func{arctanh}\sqrt{1-4\lambda ^{2}}%
}{2\sqrt{1-4\lambda ^{2}}}.  \label{1plus}
\end{equation}

Another trivial consistency check is obtained when we add to the Swanson
model a multiple of the Casimir operator $C=K_{0}^{2}-\{K_{+},K_{-}\}/2$ and
consider 
\begin{equation}
H_{C}=H_{l}+\kappa C\text{ }=(\mu _{0}+\kappa )K_{0}+\mu _{+}K_{+}+\mu
_{-}K_{-}+\kappa K_{0}^{2}-\kappa K_{+}K_{-}~\ \ \ \ \ \ \text{for }\kappa
\in \mathbb{R}.  \label{Cas}
\end{equation}
Since the Casimir operator is Hermitian and commutes with $\eta $ no further
constraint should result from this modification when compared with the
non-Hermitian linear case. In fact, the linear case together with the
constraining equations will produce the Casimir operator. Starting with the
latter case and replacing $\mu _{0}\rightarrow \mu _{0}+\kappa $, we can
interpret $\kappa =-\mu _{+-}$ according to (\ref{T1}). When $\mu _{+-}\neq
0 $ we can satisfy the constraints (\ref{C1}) by $\mu _{+-}=-\mu _{00}$ and
setting all remaining $\mu $'s with double subscripts to be zero, which is
obviously satisfied by (\ref{Cas}), together with 
\begin{equation}
\frac{\tanh 2\theta }{\theta /\varepsilon }=\frac{\mu _{-}-\mu _{+}}{\mu
_{-}+\mu _{+}-2\lambda \mu _{0}}.
\end{equation}
\qquad

We conclude this section by making use of the constraining equation (\ref{T1}%
) and re-express the operator $\eta $ in (\ref{eta}) purely as as a function
of $\lambda \in \lbrack -\frac{1}{2},\frac{1}{2}]\backslash \{0\}$ 
\begin{equation}
\eta (\lambda )=\exp {\left[ \frac{K_{0}+\lambda (K_{+}+K_{-})}{\sqrt{%
1-4\lambda ^{2}}}\func{arctanh}{F}(\lambda )\right] },  \label{eta1}
\end{equation}%
where 
\begin{equation}
F(\lambda ):=\sqrt{1-4\lambda ^{2}}\frac{\lambda (\mu _{-}-\mu _{+})+\mu
_{++}-\mu _{--}}{\lambda (\mu _{-}+\mu _{+})+2\lambda ^{2}(\mu _{+-}-\mu
_{0})-(\mu _{++}+\mu _{--})},
\end{equation}%
subject to the constraints (\ref{C1}).

\paragraph{Hermitian counterpart}

Using the explicit solution (\ref{eta1}) we can compute the Hermitian
counterpart $h_{K}$ using the formula (\ref{herm}). As expected from similar
calculations previously carried out in this context the explicit
non-Hermitian Hamiltonian turns out to be rather complicated when compared
to the fairly simple non-Hermitian Hamiltonian (\ref{HK}). Nonetheless, it
may be computed exactly and we find the coefficients in (\ref{hham}) to be
given by 
\begin{eqnarray}
\hat{\mu}_{0} &=&A_{0}+2(\mu _{-}-\mu _{+})\lambda B_{0},  \label{mu1} \\
\hat{\mu}_{+} &=&\hat{\mu}_{-}=\lambda A_{+}+\frac{1}{2}\left[ \mu _{-}-\mu
_{+}+2(\mu _{--}-\mu _{++})\lambda \right] B_{0}, \\
\hat{\mu}_{00} &=&\frac{1}{2\lambda }A_{00}+2(\mu _{--}-\mu _{++})B_{0}, \\
\hat{\mu}_{+-} &=&A_{+-}+(\mu _{--}-\mu _{++})B_{0}, \\
\hat{\mu}_{++} &=&\hat{\mu}_{--}=\lambda A_{++}+\frac{1}{2}(\mu _{--}-\mu
_{++})B_{0}, \\
\hat{\mu}_{+0} &=&\hat{\mu}_{0-}=2A_{++}+\left( \frac{1+4\lambda ^{2}}{%
2\lambda }\right) (\mu _{--}-\mu _{++})B_{0},
\end{eqnarray}%
where we further abbreviated 
\begin{eqnarray}
A_{0} &=&\mu _{0}-\frac{2\lambda }{1-4\lambda ^{2}}\frac{(\mu _{-}-\mu
_{+})(\mu _{--}+3\mu _{++}-2\lambda \mu _{+0})}{(\mu _{--}-\mu _{++})}, \\
A_{00} &=&\frac{1}{(1-4\lambda ^{2})(\mu _{--}-\mu _{++}-\lambda (\mu
_{-}-\mu _{+}))}\left[ 2(\mu _{--}\mu _{+}-\mu _{++}\mu _{-})\right. \\
&&-2\lambda (\mu _{--}-\mu _{++})(\mu _{0}+\mu _{++}+\mu _{--})+2\lambda
^{2}(\mu _{-}-\mu _{+})(\mu _{--}+\mu _{++}+\mu _{+-})+  \notag \\
&&\left. -8\lambda ^{3}(\mu _{--}-\mu _{++})(\mu _{--}+\mu _{++}-\mu
_{+-})+8\lambda ^{4}(\mu _{-}-\mu _{+})(\mu _{--}+\mu _{++}-\mu _{+-})\right]
,  \notag \\
A_{+-} &=&\frac{1}{(1-4\lambda ^{2})(\mu _{--}-\mu _{++}-\lambda (\mu
_{-}-\mu _{+}))}\left\{ (\mu _{++}-\mu _{--})(\mu _{--}+\mu _{++}-\mu
_{+-})\right. \\
&&-\lambda \left[ \mu _{+-}(\mu _{-}-\mu _{+})-(\mu _{-}+\mu _{+})(\mu
_{--}-\mu _{++})\right]  \notag \\
&&\left. -2\lambda ^{2}(\mu _{--}-\mu _{++})(\mu _{+-}+\mu _{0})+4\lambda
^{3}(\mu _{-}-\mu _{+})\mu _{+-}\right\} ,  \notag \\
A_{+} &=&\frac{1}{(1-4\lambda ^{2})\lambda (\mu _{--}-\mu _{++})}\left\{
-\lambda \left[ \mu _{--}^{2}-(\mu _{-}-\mu _{+})\mu _{+0}+2\mu _{--}\mu
_{++}-3\mu _{++}^{2}\right] \right. \\
&&\left. +\mu _{+}(\mu _{--}+\mu _{++})-2\mu _{-}\mu _{++}-2\lambda ^{2}(\mu
_{--}-\mu _{++})(\mu _{-}+\mu _{-}-\mu _{+0})\right\} ,  \notag \\
A_{++} &=&\frac{\lambda \mu _{+0}-\mu _{++}-2\lambda ^{2}(\mu _{--}+\mu
_{++})}{(1-4\lambda ^{2})\lambda }, \\
B_{0} &=&2\frac{\sqrt{2\mu _{++}(\mu _{--}+\mu _{++})-\lambda (\mu
_{--}+3\mu _{++})\mu _{+0}+\lambda ^{2}\left[ \mu _{+0}^{2}+(\mu _{--}-\mu
_{++})^{2}\right] }}{(1-4\lambda ^{2})(\mu _{--}-\mu _{++})}.  \label{mu3}
\end{eqnarray}%
Clearly this Hamiltonian does not constitute an obvious starting point,
whereas the non-Hermitian Hamiltonian $H_{J}$ is fairly simple and natural
to consider. We could also express the Hermitian version in a simple fashion
by solving (\ref{mu1})-(\ref{mu3}) for the $\mu $s, such that instead $H_{J}$
would acquires a complicated form. However, the construction procedure
itself is only meaningful in the direction $H_{J}\rightarrow h_{J}$ and not $%
h_{J}\rightarrow H_{J}$.

\subsubsection{Generic non-Hermitian non-reducible Hamiltonian\label{nonred}}

Remarkably in contrast to the previously analysed purely linear case there
exists a second non-equivalent type of solution. We find that (\ref{cc1}), (%
\ref{cc2}) and (\ref{cc3}) are also solved by the four constraints 
\begin{eqnarray}
\mu _{+}-\mu _{-} &=&2\lambda (\mu _{++}-\mu _{--}),  \label{v1} \\
\mu _{+0}-\mu _{0-} &=&2(\mu _{+}-\mu _{-}),  \label{v2} \\
\mu _{+0} &=&2\mu _{+}+2(\mu _{+-}-\mu _{0})\lambda ,  \label{v3} \\
\mu _{+} &=&\lambda (\mu _{0}+\mu _{00}+2\mu _{++})-2\lambda ^{2}(\mu
_{-}-\mu _{+}+\mu _{+0})~~,~  \label{v4}
\end{eqnarray}%
together with 
\begin{equation}
\frac{\tanh 4\theta }{\theta /\varepsilon }=\frac{\mu _{--}-\mu _{++}}{\mu
_{--}+\mu _{++}+\lambda (\mu _{+}-\mu _{-}-\mu _{+0})}.  \label{v5}
\end{equation}%
Notice that this solution can not be reduced to the cases of a non-Hermitian
linear term plus Hermitian bilinear combination or a Hermitian linear term
plus a non-Hermitian bilinear combination as discussed in sections \ref%
{NHLHB} or \ref{HLNHB}, respectively. This is seen from (\ref{v1}) and (\ref%
{v2}) as $\mu _{+}=\mu _{-}$ implies $\mu _{++}=\mu _{--}$, $\mu _{+0}=\mu
_{0-}$ and vice versa, such that it is impossible to convert one part into a
Hermitian one while keeping the other non-Hermitian.

As for the foregoing set of constraints there are some interesting subcases.
For instance, we can consider again the situation where the Hamiltonian does
not contain any generators of the type $K_{-}$, i.e. $\mu _{-}=\mu _{--}=\mu
_{0-}=0$. Then the constraints simplify to 
\begin{eqnarray}
2\lambda \mu _{++} &=&\mu _{+},\qquad \mu _{+0}=2\mu _{+},\qquad \mu
_{00}=2\lambda \mu _{+}-\mu _{0},\qquad \mu _{+-}=\mu _{0},\quad
\label{2minus} \\
\varepsilon &=&-\frac{1}{4\sqrt{1-4\lambda ^{2}}}\func{arctanh}\left( \frac{%
\sqrt{1-4\lambda ^{2}}}{1-2\lambda ^{2}}\right) .  \notag
\end{eqnarray}%
Similarly, if the Hamiltonian does not contain any generators of the type $%
K_{+}$, i.e., $\mu _{+}=\mu _{++}=\mu _{0+}=0$, the constraints reduce to 
\begin{eqnarray}
2\lambda \mu _{--} &=&\mu _{-},\qquad \mu _{0-}=2\mu _{-},\qquad \mu
_{00}=2\lambda \mu _{-}-\mu _{0},\qquad \mu _{+-}=\mu _{0}  \label{2plus} \\
\varepsilon &=&\frac{1}{4\sqrt{1-4\lambda ^{2}}}\func{arctanh}\left( \frac{%
\sqrt{1-4\lambda ^{2}}}{1-2\lambda ^{2}}\right) .  \notag
\end{eqnarray}%
Note that also for this reduced case the solutions (\ref{1minus}) and (\ref%
{2minus}) as well as (\ref{1plus}) and (\ref{2plus}) are different.

As before we can also in this case use of the constraining equation (\ref{T1}%
) and re-express the operator $\eta $ in (\ref{eta}) purely as a function of 
$\lambda \in \lbrack -\frac{1}{2},\frac{1}{2}]\backslash \{0\}$ 
\begin{equation}
\eta (\lambda )=\exp {\left[ \frac{K_{0}+\lambda (K_{+}+K_{-})}{2\sqrt{%
1-4\lambda ^{2}}}\func{arctanh}{G}(\lambda )\right] },  \label{553}
\end{equation}%
where 
\begin{equation}
G(\lambda ):=\sqrt{1-4\lambda ^{2}}\frac{(\mu _{--}-\mu _{++})}{\mu
_{--}+\mu _{++}+\lambda (\mu _{+}-\mu _{-}-\mu _{+0})},
\end{equation}%
subject to the constraints (\ref{v1}) and (\ref{v4}).

\paragraph{Hermitian counterpart}

Using again the explicit solution (\ref{eta1}) we can compute the Hermitian
counterpart $h_{K}$ using the formula (\ref{553}). Once more the expressions
are quite cumbersome

\begin{eqnarray}
\hat{\mu}_{0} &=&C_{0}+4\lambda ^{2}D_{0}, \\
\hat{\mu}_{+} &=&\hat{\mu}_{-}=C_{+}+2\lambda D_{0}, \\
\hat{\mu}_{00} &=&2(C_{00}+4\lambda ^{2}D_{0}), \\
\hat{\mu}_{+-} &=&C_{+-}+4\lambda ^{2}D_{0}, \\
\hat{\mu}_{++} &=&\hat{\mu}_{--}=\lambda C_{++}+(1-2\lambda ^{2})D_{0}, \\
\hat{\mu}_{+0} &=&\hat{\mu}_{0-}=2(C_{++}+2\lambda D_{0}),
\end{eqnarray}%
where further abbreviated 
\begin{eqnarray}
C_{0} &=&\frac{\mu _{0}-\lambda (\mu _{-}-\mu _{+})-4\lambda ^{2}(\mu
_{++}+\mu _{0})+2\lambda ^{3}(\mu _{-}+\mu _{+})+4\lambda ^{4}(\mu _{+-}-\mu
_{0})}{1-4\lambda ^{2}}, \\
C_{00} &=&\frac{\mu _{00}}{2}-\frac{2\lambda ^{2}\left[ \mu _{--}+\mu
_{++}-\lambda \mu _{+0}-2\lambda ^{2}(\mu _{--}-\mu _{++})\right] }{%
1-4\lambda ^{2}}, \\
C_{+-} &=&C_{0}+\mu _{+-}-\mu _{0}, \\
C_{+} &=&\frac{\mu _{+}-2\lambda \mu _{++}-\lambda ^{2}(\mu _{-}+\mu
_{+})+2\lambda ^{3}(\mu _{+-}-\mu _{0})}{1-4\lambda ^{2}}, \\
C_{++} &=&\frac{\mu _{+0}-4\lambda \mu _{++}-2\lambda ^{2}\mu _{+0}-4\lambda
^{3}(\mu _{--}-\mu _{++})}{2(1-4\lambda ^{2})}, \\
D_{0} &=&\frac{1}{2\,\,\left( 4\,{\lambda }^{2}-1\right) }\left\{ 4\,{\mu
_{--}}\,{\mu _{++}+{\lambda }^{2}\,\left[ {{\mu _{+0}^{2}}}+8\,{\mu _{++}}%
\,\left( {\mu _{++}}-{\mu _{--}}\right) \right] }\right. \\
&&\left. {-2\,{\lambda }\,{\mu _{+0}}\,\left( {\mu _{--}}+{\mu _{++}}\right)
+}4\,{\lambda }^{3}\,{\mu _{+0}}\,\left( {\mu _{--}}-{\mu _{++}}\right) +4\,{%
\lambda }^{4}\,{\left( {\mu _{--}}-{\mu _{++}}\right) }^{2}\right\} ^{1/2}. 
\notag
\end{eqnarray}%
Again this demonstrates the general feature that some fairly simple
non-Hermitian Hamiltonians possess quite complicated isospectral Hermitian
counterparts.

\subsubsection{A simpler metric, the case $\protect\lambda =0$}

In the previous discussion we have excluded the case $\lambda =0$, which
equals $\nu =0$ in our ansatz for the metric (\ref{eta}). This case may be
dealt with separately and in fact is fairly easy, as $\eta $ simplifies
considerably because it only depends on the generator $K_{0}$. In this
situation also the constraints turn out to be far simpler

\begin{equation}
\mu _{--}\mu _{+}^{2}=\mu _{++}\mu _{-}^{2},\qquad \mu _{--}\mu
_{+0}^{2}=\mu _{++}\mu _{0-}^{2}\qquad \text{and\qquad }\varepsilon =\frac{1%
}{8}\ln \frac{\mu _{--}}{\mu _{++}}
\end{equation}%
and even the Hermitian counterpart Hamiltonian becomes fairly compact too 
\begin{eqnarray}
h_{\varepsilon } &=&\mu _{0}K_{0}+\mu _{+}e^{2\varepsilon }(K_{+}+K_{-})+\mu
_{00}K_{0}^{2}+\mu _{+-}K_{+}K_{-}+\mu _{++}e^{4\varepsilon
}(K_{+}^{2}+K_{-}^{2}) \\
&&+\mu _{+0}e^{2\varepsilon }(K_{+}K_{0}+K_{0}K_{-}).  \notag
\end{eqnarray}%
This suggests that the simple metric $\eta =e^{2\varepsilon K_{0}}$ may be
employed as an easy transformation also for other more complicated
Hamiltonians.

\subsubsection{Two further simple cases $\protect\lambda =\pm 1/2$}

Finally let us also investigate the other boundary values for the parameter $%
\lambda $, that is $\lambda =\pm 1/2$. \ In this case the constraints are

\begin{eqnarray}
\mu _{++} &=&\pm (\mu _{+}-2\mu _{-})+\frac{(\mu _{-}-\mu _{+})\mu _{+-}}{%
\mu _{+0}-\mu _{0-}+2(\mu _{-}-\mu _{+})}\pm \frac{(\mu _{-}-\mu _{+})(\mu
_{0-}-2\mu _{-}\pm \mu _{0})}{\mu _{0-}-\mu _{+0}-2(\mu _{-}-\mu _{+})}~~~~
\\
&&+\frac{\mu _{0}+\mu _{00}}{2}+\frac{(\mu _{0-}-\mu _{+0}-2(\mu _{-}-\mu
_{+}))(\mu _{+0}-2(\mu _{0}+\mu _{00}-2\mu _{-}))}{4(\mu _{0-}-2\mu _{-}\pm
(\mu _{0}-\mu _{+-}))},  \notag \\
\mu _{--} &=&\mp \mu _{+}+\frac{(\mu _{-}-\mu _{+})\mu _{+-}}{\mu _{+0}-\mu
_{0-}+2(\mu _{-}-\mu _{+})}+\frac{(\mu _{-}-\mu _{+})(\mu _{0}+\mu
_{0-}-2\mu _{-})}{\mu _{0-}-\mu _{+0}-2(\mu _{-}-\mu _{+})} \\
&&+\frac{\mu _{0}+\mu _{00}}{2}+\frac{(2(\mu _{0}+\mu _{00})\mp (\mu
_{0-}+4\mu _{+}))(\mu _{0-}-\mu _{0+}+2(\mu _{+}-\mu _{-}))}{4(\mu
_{+0}-2\mu _{+}\pm (\mu _{0}-\mu _{+-}))},  \notag \\
\varepsilon &=&\frac{\mu _{0-}-\mu _{+0}-2(\mu _{-}-\mu _{+})}{2(\mu
_{0-}+\mu _{+0}-2(\mu _{+}+\mu _{-})\pm 2(\mu _{0}-\mu _{+-}))}.
\end{eqnarray}%
The general Hermitian counterpart turns out to have a very complicated form,
but there are some simple special cases, such as

\begin{equation}
H_{\frac{1}{2}}=K_{+}-K_{-}-K_{0}+K_{0}^{2}+K_{+}K_{-}+K_{+}K_{0}+K_{0}K_{-}+%
\frac{11}{2}K_{+}^{2}+\frac{1}{2}K_{-}^{2}
\end{equation}
which is mapped into the Hermitian form

\begin{equation}
h_{\frac{1}{2}}=\mp \frac{13}{16}(K_{+}+K_{-})-\frac{23}{16}K_{0}+\frac{5}{8}%
K_{0}^{2}+\frac{13}{16}K_{+}K_{-}\mp \frac{11}{8}(K_{+}K_{0}+K_{0}K_{-})+%
\frac{61}{32}(K_{+}^{2}+K_{-}^{2})
\end{equation}%
with $\varepsilon =\mp \frac{1}{4}$ for $\lambda =\pm \frac{1}{2}$. \ 

\section{Generalised Bogoliubov transformation\label{Bogo}}

Bogoliubov transformations were first introduced with the purpose to
understand the pairing interaction in superconductivity \cite%
{Bogolyubov:1958km} and have been generalized thereafter in many different
ways, as for instance in \cite{GenBog}. In the present context they have
been applied by Swanson \cite{Swanson} as an alternative method to establish
the reality of the spectrum of a non-Hermitian Hamiltonian. Instead of
constructing an explicit similarity transformation one can make a
constraining assumption about the form of its Hermitian counterpart. The
simplest assumption to make is that the counterpart is of a harmonic
oscillator type. We will now demonstrate how the Hamiltonian $H_{K}$ can be
transformed into such a form by means of a generalized Bogoliubov
transformation. Following \cite{Swanson}, we define for this purpose two new
operators $c$ and $d$ via 
\begin{equation}
\left( 
\begin{array}{c}
d \\ 
c%
\end{array}%
\right) =\left( 
\begin{array}{cc}
\beta & -\delta \\ 
-\alpha & \gamma%
\end{array}%
\right) \left( 
\begin{array}{c}
a \\ 
a^{\dag }%
\end{array}%
\right) \qquad \text{with }\alpha ,\beta ,\gamma ,\delta \in \mathbb{C}.
\label{cd}
\end{equation}%
\qquad Demanding that these operators commute in the same manner as the
annihilation and creation operators $a,a^{\dagger }$, i.e. $[d,c]=1$, and
that they may be reduced to the former in a well defined limit yields the
constraints 
\begin{equation}
\beta \gamma -\alpha \delta =1\qquad \text{and\qquad }\beta ,\gamma \neq 0,
\label{beta}
\end{equation}%
on the complex parameters $\alpha ,\beta ,\gamma ,\delta $. Note that we do
not require the same Hermiticity conditions as for the conventional
operators $a=(a^{\dagger })^{\dagger }$, that is in general we have $c\neq
d^{\dagger }$. For our purposes we also require a definite behaviour under
the $\mathcal{PT}$-transformation. Noting that $\mathcal{PT}$: $a,a^{\dagger
}\rightarrow -a,-a^{\dagger }$ implies that $\alpha ,\beta ,\gamma ,\delta
\in \mathbb{R}$ or $i\mathbb{R}$, such that $\mathcal{PT}$: $c,d\rightarrow
-c,-d$ or $c,d$, respectively. In fact, we shall see below that demanding
unbroken $\mathcal{PT}$-symmetry requires the cofficients $\alpha ,\beta
,\gamma ,\delta $ to be purely complex. \ 

We may now simply invert the relations (\ref{cd})

\begin{equation}
\left( 
\begin{array}{c}
a \\ 
a^{\dag }%
\end{array}%
\right) =\left( 
\begin{array}{cc}
\gamma & \delta \\ 
\alpha & \beta%
\end{array}%
\right) \left( 
\begin{array}{c}
d \\ 
c%
\end{array}%
\right)  \label{cdinv}
\end{equation}%
and express the generators $K_{0},K_{\pm }$ in the concrete two-boson
representation (\ref{bosrep}) in terms of these new operators 
\begin{eqnarray}
K_{0} &=&\frac{1}{2}(\gamma \beta +\delta \alpha )cd+\frac{1}{2}\delta \beta
c^{2}+\frac{1}{2}\gamma \alpha d^{2}+\frac{1}{2}\delta \alpha +\frac{1}{4},
\label{k1} \\
K_{+} &=&\alpha \beta (cd+\frac{1}{2})+\frac{1}{2}\beta ^{2}c^{2}+\frac{1}{2}%
\alpha ^{2}d^{2},  \label{k2} \\
K_{-} &=&\gamma \delta (cd+\frac{1}{2})+\frac{1}{2}\delta ^{2}c^{2}+\frac{1}{%
2}\gamma ^{2}d^{2},  \label{k3}
\end{eqnarray}%
which by construction satisfy the same $su(1,1)$-commutation relations (\ref%
{comK}). Naturally we can now define the analogues of the generators $%
K_{0},K_{\pm }$ in terms of the operators $c,d$ 
\begin{equation}
\check{K}_{0}=\frac{1}{2}(cd+\frac{1}{2}),\qquad \qquad \check{K}_{+}=\frac{1%
}{2}cc,\qquad \qquad \check{K}_{-}=\frac{1}{2}dd,
\end{equation}%
such that 
\begin{equation}
\left( 
\begin{array}{c}
K_{0} \\ 
K_{+} \\ 
K_{-}%
\end{array}%
\right) =\left( 
\begin{array}{ccc}
\gamma \beta +\delta \alpha & \beta \delta & \alpha \gamma \\ 
2\alpha \beta & \beta ^{2} & \alpha ^{2} \\ 
2\gamma \delta & \delta ^{2} & \gamma ^{2}%
\end{array}%
\right) \left( 
\begin{array}{c}
\check{K}_{0} \\ 
\check{K}_{+} \\ 
\check{K}_{-}%
\end{array}%
\right) .  \label{KK}
\end{equation}%
Inverting the relation (\ref{KK}), upon using (\ref{beta}) we may also
express the $\check{K}_{0},\check{K}_{\pm }$ in terms of $K_{0},K_{\pm }$ 
\begin{equation}
\left( 
\begin{array}{c}
\check{K}_{0} \\ 
\check{K}_{+} \\ 
\check{K}_{-}%
\end{array}%
\right) =\left( 
\begin{array}{ccc}
\gamma \beta +\delta \alpha & -\gamma \delta & -\alpha \beta \\ 
-2\gamma \alpha & \gamma ^{2} & \alpha ^{2} \\ 
-2\delta \beta & \delta ^{2} & \beta ^{2}%
\end{array}%
\right) \left( 
\begin{array}{c}
K_{0} \\ 
K_{+} \\ 
K_{-}%
\end{array}%
\right) .
\end{equation}%
Replacing in $H_{K}$ the generators $K_{0},K_{\pm }$ by the newly defined
generators $\check{K}_{0},\check{K}_{\pm }$ we can transform the Hamiltonian
into the form 
\begin{equation}
H_{K}=\dsum\limits_{l=0,\pm }\check{\mu}_{l}\check{K}_{l}+\dsum%
\limits_{n,m=0,\pm }\check{\mu}_{nm}:\check{K}_{n}\check{K}_{m}:.
\label{HKhat}
\end{equation}%
Notice that due to the identity $8\check{K}_{+}\check{K}_{-}=8\check{K}_{0}%
\check{K}_{0}-8\check{K}_{0}+1$ not all coefficients $\check{\mu}_{l}$, $%
\check{\mu}_{nm}$ are uniquely defined. However, this ambiguity will not
play any role in our analysis as the relevant equations will be insensitive
to these redefinitions. Demanding that the Hamiltonian in terms of the new
generators $\check{K}_{0}$, $\check{K}_{\pm }$ acquires the form of a
harmonic oscillator plus a Casimir operator means we have to set the
constants $\hat{\mu}_{+}$, $\hat{\mu}_{-}$, $\hat{\mu}_{++}$, $\hat{\mu}%
_{--} $, $\hat{\mu}_{+0}$, $\hat{\mu}_{0-}$ to zero. Expressing these
constraints through the original constants in (\ref{HK}) yields the
equations 
\begin{eqnarray}
\mu _{++}y^{4}+\mu _{+0}y^{3}+(\mu _{+-}+\mu _{00})y^{2}+\mu _{0-}y+\mu
_{--} &=&0,  \label{cong} \\
\mu _{--}z^{4}+\mu _{0-}z^{3}+(\mu _{+-}+\mu _{00})z^{2}+\mu _{+0}z+\mu
_{++} &=&0,  \notag
\end{eqnarray}%
\begin{equation*}
\mu _{+0}y^{3}z+4\mu _{++}y^{3}+2(\mu _{+-}+\mu _{00})y^{2}z+3\mu
_{+0}y^{2}+3\mu _{0-}yz+2(\mu _{+-}+\mu _{00})y+4\mu _{--}z+\mu _{0-}=0,
\end{equation*}%
\begin{equation*}
\mu _{0-}yz^{3}+4\mu _{--}z^{3}+2(\mu _{+-}+\mu _{00})yz^{2}+3\mu
_{0-}z^{2}+3\mu _{+0}yz+2(\mu _{+-}+\mu _{00})z+4\mu _{++}y+\mu _{+0}=0,
\end{equation*}%
\begin{eqnarray*}
(\mu _{0-}-\mu _{-})yz^{3}+(2\mu _{+-}+\mu _{00}-\mu _{0})yz^{2}+2\mu
_{--}z^{3}+(2\mu _{+0}-\mu _{+})yz+(\mu _{-}+\mu _{0-})z^{2} && \\
+(\mu _{0}+\mu _{00})z+2\mu _{++}y+\mu _{+} &=&0 \\
(\mu _{+0}-\mu _{+})y^{3}z+(2\mu _{+-}+\mu _{00}-\mu _{0})y^{2}z+2\mu
_{++}y^{3}+(2\mu _{0-}-\mu _{-})yz+(\mu _{+}+\mu _{+0})y^{2} && \\
+(\mu _{0}+\mu _{00})y+2\mu _{--}z+\mu _{-} &=&0
\end{eqnarray*}%
where we abbreviated 
\begin{equation}
y=\frac{\alpha }{\gamma }\qquad \text{and}\qquad z=\frac{\delta }{\beta }.
\end{equation}%
We will now systematically solve the six equations (\ref{cong}). When $%
\alpha $, $\delta \neq 0$ the equations reduce to the simpler form 
\begin{eqnarray}
z^{2}(\mu _{00}+\mu _{+-}) &=&\mu _{++}(1+4yz+y^{2}z^{2}),  \label{s1} \\
z^{2}(\mu _{+-}-\mu _{0}) &=&\mu _{++}(1+yz)^{2}+\mu _{+}z^{3}+\mu _{-}z,
\label{s2} \\
\mu _{--}z^{2} &=&\mu _{++}y^{2},  \label{s3} \\
\mu _{+0}z &=&-2\mu _{++}(1+yz),  \label{s4} \\
\mu _{-}z &=&\mu _{+}y,  \label{s5} \\
\mu _{0-}z &=&\mu _{+0}y.  \label{s6}
\end{eqnarray}%
Similarly as in section 3 the solutions fall into different classes
distinguished by vanishing linear or bilinear combinations.

\subsection{Genuinely non-Hermitian non-reducible Hamiltonian}

We start to solve the six constraints (\ref{s1})-(\ref{s6}) for the generic
case by demanding $\mu _{+},\mu _{-}\neq 0$ and $\mu _{++},\mu _{--},\mu
_{+0},\mu _{0-}\neq 0$. We find the unique solution 
\begin{eqnarray}
\mu _{-} &=&\frac{y}{z}\mu _{+},\qquad \mu _{--}=\frac{\mu _{-}^{2}}{\mu
_{+}^{2}}\mu _{++},\qquad \mu _{0-}=\frac{\mu _{-}}{\mu _{+}}\mu
_{+0},\qquad y=\frac{\pm \vartheta -\mu _{+0}/4}{\mu _{++}},  \label{non1} \\
\mu _{+-} &=&\mu _{0}-\frac{\mu _{+}\mu _{+0}}{2\mu _{++}}+\frac{\mu
_{+0}^{2}}{4\mu _{++}},\qquad \mu _{00}=-\mu _{0}+\frac{\mu _{+}\mu _{+0}}{%
2\mu _{++}}+\frac{2\mu _{-}\mu _{++}}{\mu _{+}},  \label{non2}
\end{eqnarray}%
with the abbreviation $\vartheta :=\sqrt{\mu _{+0}^{2}/16-\mu _{++}^{2}\mu
_{+}/\mu _{-}}$. The Hamiltonian $H_{K}$\ in (\ref{HK}) or in other words
the Hamiltonians $H_{\check{K}}$ in (\ref{HKhat}) can now be expressed
entirely in terms of the number operator $\check{N}=cd$ and acquires the
simple form 
\begin{equation}
H_{K}\ =\frac{\vartheta ^{2}}{\mu _{++}}(\check{N}^{2}+\check{N})\pm \frac{%
\vartheta (\mu _{+0}-2\mu _{+})}{2\mu _{++}}\left( \check{N}+\frac{1}{2}%
\right) +\frac{3\mu _{0}}{16}-\frac{3\mu _{+}\mu _{+0}}{32\mu _{++}}+\frac{%
\mu _{+0}^{2}}{16\mu _{++}}-\frac{5\mu _{-}\mu _{++}}{8\mu _{+}}.
\end{equation}%
In analogy to the harmonic oscillator case we may now easily construct the
eigensystem for this Hamiltonian. Defining the states $|\check{n}\rangle
=(n!)^{-1/2}c^{n}|\check{0}\rangle $ with $d|\check{0}\rangle =0$ we have $%
\check{N}|\check{n}\rangle =n|\check{n}\rangle $. Note that demanding $%
\mathcal{PT}$-symmetry for the states $|\check{n}\rangle $ requires that $%
\mathcal{PT}$: $c\rightarrow c$, which in turn implies $\alpha ,\beta
,\gamma ,\delta \in i\mathbb{R}$. Demanding that the eigenspectrum is real
and bounded from below imposes the further constraints

\begin{equation}
\mu _{++}>0\qquad \text{and\qquad }\mu _{-}\mu _{+0}^{2}>16\mu _{++}^{2}\mu
_{+}.  \label{b1}
\end{equation}

It is now interesting to compare this result with our previous construction
for the isospectral counterpart in section \ref{nonred}. Using the
constraints (\ref{non1}) and (\ref{non2}) we may solve the conditions (\ref%
{v1})-(\ref{v4}) for the similarity transformation needed to be able to
construct a well defined metric operator by

\begin{equation}
\mu _{+0}=2\mu _{+}\qquad \text{and\quad }\lambda =\frac{\mu _{+}^{2}}{2(\mu
_{+}+\mu _{-})\mu _{++}},
\end{equation}%
such that (\ref{v5}) acquires the form 
\begin{equation}
\frac{\tanh {4\theta }}{\theta /\varepsilon }=\frac{2(\mu _{-}^{2}-\mu
_{+}^{2})\mu _{++}^{2}}{2(\mu _{-}^{2}+\mu _{+}^{2})\mu _{++}^{2}-\mu
_{+}^{4}}.
\end{equation}%
Thus upon these constraints, the two constructions coincide, if besides (\ref%
{b1}) we also demand that $\mu _{+}^{2}\leq \mu _{++}\left\vert \mu _{+}+\mu
_{-}\right\vert $ since $\left\vert \lambda \right\vert \leq 1/2$. This
means that in this situation we do not only have an explicit similarity
transformation, a well defined metric and a Hermitian counterpart, but in
addition we know the exact eigenspectrum and eigenfunctions. Relaxing these
conditions it also implies that there must be a larger class of similarity
transformations not covered by the ansatz (\ref{eta}) for the operator $\eta 
$. As already mentioned we might be loosing out on some possibilities by
demanding $\eta $ to be Hermitian. A further natural generalisation would be
to include also bilinear combinations into the argument of the exponential
in the expression for $\eta $.

\subsection{Hermitian linear term and non-Hermitian bilinear combinations}

It seems natural that we mimic the same cases as for the construction of the
metric in section \ref{met}. However, when tuning the linear term to be
Hermitian by demanding $\mu _{+}=\mu _{-}$ the constraints (\ref{s3}), (\ref%
{s5}) and (\ref{s6}) imply that $\mu _{++}=\mu _{--}$ and $\mu _{+0}=\mu
_{0-}$, such that also the terms involving bilinear combinations becomes
Hermitian. The case $\mu _{+}=\mu _{-}=0$ is special since the last equation
in (\ref{non2}) yields $\mu _{+}/\mu _{-}=$ $\left( \mu _{0}+\mu
_{00}\right) /(2\mu _{++})$. Using this and still demanding that $\mu
_{++},\mu _{--},\mu _{+0},\mu _{0-}\neq 0$, the solutions to (\ref{s1})-(\ref%
{s6}) become 
\begin{eqnarray}
\mu _{--} &=&\frac{\left( \mu _{0}+\mu _{00}\right) ^{2}}{4\mu _{++}},\qquad
\mu _{0-}=\frac{\mu _{+0}}{2\mu _{++}}\left( \mu _{0}+\mu _{00}\right)
,\qquad y=\frac{\pm \bar{\vartheta}-\mu _{+0}/4}{\mu _{++}},  \label{l1} \\
\mu _{+-} &=&\mu _{0}+\frac{\mu _{+0}^{2}}{4\mu _{++}},\qquad z=y\frac{2\mu
_{++}}{\mu _{0}+\mu _{00}},  \label{l2}
\end{eqnarray}%
with the abbreviation $\bar{\vartheta}:=\sqrt{\mu _{+0}^{2}/16-\mu
_{++}\left( \mu _{0}+\mu _{00}\right) /2}$. The Hamiltonian $H_{K}$\ in (\ref%
{HK}), (\ref{HKhat}) can be expressed again entirely in terms of the number
operator and acquires the simple form 
\begin{equation}
H_{K}\ =\frac{\bar{\vartheta}^{2}}{\mu _{++}}(\check{N}^{2}+\check{N})\pm 
\frac{\bar{\vartheta}\mu _{+0}}{2\mu _{++}}\left( \check{N}+\frac{1}{2}%
\right) +\frac{\mu _{+0}^{2}}{16\mu _{++}}-\frac{5}{16}\mu _{00}-\frac{\mu
_{0}}{8}.
\end{equation}%
The requirement that the eigenspectrum is real and bounded from below yields
in this case the additional constraints 
\begin{equation}
\mu _{++}>0\qquad \text{and\qquad }\mu _{+0}^{2}>8\mu _{++}\left( \mu
_{0}+\mu _{00}\right) .
\end{equation}

Interestingly when demanding (\ref{l1}) and (\ref{l2}), we can not solve the
constraints in section \ref{met} and therefore can not construct a metric
with the ansatz (\ref{eta}) in this case.

\subsection{Non-Hermitian linear case and Hermitian bilinear combinations}

Reversing the setting of the previous section we may now demand the bilinear
combinations to be Hermitian, $\mu _{++}=\mu _{--}$ and $\mu _{+0}=\mu _{0-}$%
. This is equally pathological as now the linear term becomes also Hermitian
by (\ref{s3}), (\ref{s5}) and (\ref{s6}). Nonetheless, a non-trivial limit
is obtained with $\mu _{++}=\mu _{--}=\mu _{+0}=\mu _{0-}=0$ and requiring $%
\mu _{+},\mu _{-}\neq 0$. We may then solve (\ref{s1})-(\ref{s6}) by

\begin{equation}
\mu _{-}=\frac{y}{z}\mu _{+},\qquad \mu _{+-}=-\mu _{00},\qquad y=\frac{\pm 
\tilde{\vartheta}-(\mu _{0}+\mu _{00})/2}{\mu _{+}},  \label{xx}
\end{equation}%
with the abbreviation $\tilde{\vartheta}:=\sqrt{(\mu _{0}+\mu
_{00})^{2}/4-\mu _{+}\mu _{-}}$. Once again the Hamiltonian $H_{K}$\ in (\ref%
{HK}), (\ref{HKhat}) can be expressed entirely in terms of the number
operator simplifying it to 
\begin{equation}
H_{K}\ =\pm \tilde{\vartheta}\left( \check{N}+\frac{1}{2}\right) -\frac{3\mu
_{00}}{16}.  \label{qq}
\end{equation}%
The eigenspectrum is real and bounded from below when we discard the minus
sign in (\ref{qq}) and impose the condition 
\begin{equation}
(\mu _{0}+\mu _{00})^{2}>4\mu _{+}\mu _{-}.
\end{equation}%
When setting $\mu _{00}=\mu _{+-}=0$ these expressions reduce precisely to
those found in \cite{Swanson} for the purely linear case.

Comparing now with the construction in section \ref{nonred}, we find that (%
\ref{313}) is solved by the conditions (\ref{xx}), if we further demand that 
\begin{equation}
\mu _{00}+\mu _{+-}=0,
\end{equation}%
such that (\ref{321}) becomes 
\begin{equation}
\frac{\tanh {2\theta }}{\theta /\varepsilon }=\frac{\mu _{-}-\mu _{+}}{\mu
_{-}+\mu _{+}-2\lambda (\mu _{00}+\mu _{0})}.
\end{equation}

We may also put further restrictions on the generalized Bogoliubov
transformation (\ref{cd}) itself by setting some of the constants to zero.

\subsection{Asymmetric generalized Bogoliubov transformation with $\protect%
\delta =0$}

Let us now set the $\alpha $ in (\ref{cd}) to zero. Then the equations (\ref%
{cong}) are solved by 
\begin{equation}
\ \ \mu _{+}=\mu _{++}=\mu _{+0}=0,\quad \mu _{0-}=-\frac{2\mu _{--}}{y}%
,\quad \mu _{00}=-\mu _{0}-\frac{\mu _{-}}{y},\quad \mu _{+-}=\frac{\mu _{--}%
}{y^{2}}-\mu _{00}.
\end{equation}%
In this situation the transformed Hamiltonian $H_{K}$\ (\ref{HKhat}) can be
expressed as 
\begin{equation}
H_{K}\ =\frac{\mu _{0-}^{2}}{16\mu _{--}}(\check{N}^{2}-\check{N})+\frac{\mu
_{0-}\mu _{-}}{4\mu _{--}}\left( \check{N}+\frac{1}{8}\right) +\frac{3\mu
_{0}}{16}.
\end{equation}%
Once again we may compare with the construction in section \ref{nonred}. The
operator $\eta $ can be constructed when we demand%
\begin{equation}
\mu _{+-}=\mu _{0}\qquad \text{and\qquad }\lambda =-\frac{1}{2y}
\end{equation}%
together with 
\begin{equation}
\varepsilon =\frac{1}{4\sqrt{1-\frac{1}{y^{2}}}}\mathrm{ArcTanh}\left( \frac{%
2y^{2}\sqrt{1-\frac{1}{y^{2}}}}{1-2y^{2}}\right) .
\end{equation}%
The meaningful interval $\lambda \in \lbrack -1/2,1/2]/\{0\}$ is now
translated into the condition $y\in \lbrack -1,1]/\{0\}$.

\subsection{Asymmetric generalized Bogoliubov transformation with $\protect%
\alpha =0$}

We may also put further constraints on the transformation (\ref{cd}) itself.
Then the equations (\ref{cong}) are solved by 
\begin{equation}
\ \ \mu _{-}=\mu _{--}=\mu _{0-}=0,\quad \mu _{+0}=-\frac{2\mu _{++}}{z}%
,\quad \mu _{00}=-\mu _{0}-\frac{\mu _{+}}{z},\quad \mu _{+-}=\frac{\mu _{++}%
}{z^{2}}-\mu _{00}.
\end{equation}%
Now the transformed Hamiltonian $H_{K}$\ (\ref{HKhat}) can be expressed as 
\begin{equation}
H_{K}\ =\frac{\mu _{+0}^{2}}{16\mu _{++}}(\check{N}^{2}-\check{N})+\frac{\mu
_{+0}\mu _{+}}{4\mu _{++}}\left( \check{N}+\frac{1}{8}\right) +\frac{3\mu
_{0}}{16}.
\end{equation}%
The comparison with the construction in section \ref{nonred} yields now that
the operator $\eta $ can be constructed when we demand%
\begin{equation}
\mu _{+-}=\mu _{0}\qquad \text{and\qquad }\lambda =-\frac{1}{2z}
\end{equation}%
together with 
\begin{equation}
\varepsilon =\frac{1}{4\sqrt{1-\frac{1}{z}}}\mathrm{ArcTanh}\left( \frac{%
2z^{2}\sqrt{1-\frac{1}{z^{2}}}}{1-2z^{2}}\right)
\end{equation}%
Now $\lambda \in \lbrack -1/2,1/2]/\{0\}$ is translated to the condition $%
z\in \lbrack -1,1]/\{0\}$.

As a trivial consistency we observe that for $\alpha =\delta =0$, i.e. when $%
y=z=0$, the transformation (\ref{cd}) becomes the identity and we have the
vanishing of all coefficients except for $\mu _{0}$, $\mu _{00}$ and $\mu
_{+-}$. Thus the initial Hamiltonian is already Hermitian and just
corresponds to the harmonic oscillator displaced by a Casimir operator. The
configuration when the constants $\mu _{+}$, $\mu _{-}$,$\,\mu _{++}$, $\mu
_{--}$, $\mu _{+0}${\normalsize \ }and{\normalsize \ $\mu _{0-}$ }vanish is
obviously of little interest.

For completeness we also comment on the case {\normalsize $yz=-1$} for which
we may also find an explicit solution. However, in this situation the
coefficients in front of $\check{K}_{0}^{2}$ and $\check{K}_{0}$ are not
positive and consequently this scenario is of little physical relevance.

\section{Some concrete realisations}

Let us finish our generic discussion with a few comments related to some
concrete realisations of the algebras discussed. The most familiar
representation of the $su(1,1)$ is probably the aforementioned two-boson
representation (\ref{bosrep}), but also the realisation for $n=1$ in (\ref%
{bosrep}) plays an important role for instance in the study of the
Jaynes-Cummings model \cite{KulishJay}. With the usual identifications for
the creation and annihilation operators in terms of differential operators
in x-space it is then straightforward to express $H_{K}$ in terms maximally
quartic in the position and momentum operators, albeit not in its most
general form, 
\begin{equation}
H_{xp}=\gamma _{0}+\gamma _{1}\hat{x}^{2}+\gamma _{2}\hat{p}^{2}+\gamma _{3}%
\hat{x}^{4}+\gamma _{4}\hat{p}^{4}+\imath \gamma _{5}\hat{x}\hat{p}+\gamma
_{6}\hat{x}^{2}\hat{p}^{2}+\imath \gamma _{7}\hat{x}\hat{p}^{3}+\imath
\gamma _{8}\hat{x}^{3}\hat{p}.  \label{Hg}
\end{equation}%
The coefficients $\gamma _{i}$ in (\ref{Hg}) and the $\mu _{l}$, $\mu _{n,m}$
in (\ref{HK}) are related as 
\begin{equation}
\left( 
\begin{array}{l}
\gamma _{0} \\ 
\gamma _{1} \\ 
\gamma _{2} \\ 
\gamma _{3} \\ 
\gamma _{4} \\ 
\gamma _{5} \\ 
\gamma _{6} \\ 
\gamma _{7} \\ 
\gamma _{8}%
\end{array}%
\right) =\frac{1}{16}\left( 
\begin{array}{rrrrrrrrr}
0 & -4 & 4 & -2 & 3 & 3 & 1 & 2 & -2 \\ 
4 & 4 & 4 & 0 & -6 & 6 & -4 & -5 & 1 \\ 
4 & -4 & -4 & 0 & 6 & -6 & -4 & -1 & 5 \\ 
0 & 0 & 0 & 1 & 1 & 1 & 1 & 1 & 1 \\ 
0 & 0 & 0 & 1 & 1 & 1 & 1 & -1 & -1 \\ 
0 & -8 & 8 & -4 & 12 & 12 & -4 & 4 & -4 \\ 
0 & 0 & 0 & 2 & -6 & -6 & 2 & 0 & 0 \\ 
0 & 0 & 0 & 0 & 4 & -4 & 0 & -2 & 2 \\ 
0 & 0 & 0 & 0 & -4 & 4 & 0 & -2 & 2%
\end{array}%
\right) \left( 
\begin{array}{l}
\mu _{0} \\ 
\mu _{+} \\ 
\mu _{-} \\ 
\mu _{00} \\ 
\mu _{++} \\ 
\mu _{--} \\ 
\mu _{+-} \\ 
\mu _{+0} \\ 
\mu _{0-}%
\end{array}%
\right) .  \label{zz}
\end{equation}%
Since the determinant of the matrix in (\ref{zz}) is non-vanishing we may
also express the $\mu _{l}$, $\mu _{n,m}$ in terms of the $\gamma _{i}$,
which then translates the constraining equations and the coefficient
occurring in the Hermitian counterparts too. It is interesting to note that
the argument $2\varepsilon \lbrack K_{0}+\lambda (K_{+}+K_{-})]$ in the
exponential of the operator $\eta $ becomes $\varepsilon \lbrack \frac{1}{2}(%
\hat{x}^{2}+\hat{p}^{2})+\lambda (\hat{x}^{2}-\hat{p}^{2})+1]$,\ such that
at the boundaries of the interval in which $\lambda $ takes its values $%
\lambda =1/2$ and $\lambda =-1/2$ the operator and therefore the metric
becomes a function only of $\hat{x}$ and $\hat{p}$, respectively.

There are plenty of other representations. An interesting one is for
instance one mentinoned in \cite{Quesne} 
\begin{eqnarray}
K_{0} &=&\frac{1}{4\xi }\left( -\frac{d^{2}}{dr^{2}}+\frac{g}{r^{2}}+\xi
^{2}r^{2}\right) ,  \label{q2} \\
K_{\pm } &=&\frac{1}{4\xi }\left( \frac{d^{2}}{dr^{2}}-\frac{g}{r^{2}}+\xi
^{2}r^{2}\mp \xi \left( 2r\frac{d}{dr}+1\right) \right) ,  \label{q3}
\end{eqnarray}%
which may also be used to relate to Calogero models \cite{Cal1}. Using this
representation $H_{K}$ may be expressed as a differential operator in $r$ 
\begin{eqnarray}
H_{R} &=&\rho _{0}+\rho _{1}\frac{d^{4}}{dr^{4}}+\rho _{2}r\frac{d^{3}}{%
dr^{3}}+\rho _{3}r^{2}\frac{d^{2}}{dr^{2}}+\rho _{4}\frac{d^{2}}{dr^{2}}%
+\rho _{5}\frac{1}{r^{2}}\frac{d^{2}}{dr^{2}}+\rho _{6}r^{3}\frac{d}{dr}+ \\
&&+\rho _{7}r\frac{d}{dr}+\rho _{8}\frac{1}{r}\frac{d}{dr}+\rho _{9}\frac{1}{%
r^{3}}\frac{d}{dr}+\rho _{10}r^{4}\frac{d}{dr}+\rho _{11}r^{2}+\rho _{12}%
\frac{1}{r^{2}}+\rho _{13}\frac{1}{r^{4}}  \notag
\end{eqnarray}%
and its corresponding Hermitian counterpart, using the constraints (\ref{C1}%
), (\ref{C2}), (\ref{C3}), (\ref{T1}), is given by 
\begin{eqnarray}
h_{R} &=&\tilde{\rho}_{0}+\tilde{\rho}_{1}\frac{d^{4}}{dr^{4}}+\tilde{\rho}%
_{3}r^{2}\frac{d^{2}}{dr^{2}}+\tilde{\rho}_{4}\frac{d^{2}}{dr^{2}}+\tilde{%
\rho}_{5}\frac{1}{r^{2}}\frac{d^{2}}{dr^{2}}+ \\
&&+\tilde{\rho}_{7}r\frac{d}{dr}+\tilde{\rho}_{9}\frac{1}{r^{3}}\frac{d}{dr}+%
\tilde{\rho}_{10}r^{4}\frac{d}{dr}+\tilde{\rho}_{11}r^{2}+\tilde{\rho}_{12}%
\frac{1}{r^{2}}+\tilde{\rho}_{13}\frac{1}{r^{4}}  \notag
\end{eqnarray}%
The $\rho $s and $\tilde{\rho}$s may be computed explicitly, but this is not
relevant for our purposes here. Keeping only linear terms in $K$ in the
Hamiltonian, we obtain 
\begin{equation}
H=\frac{(\mu _{0}-\mu _{+}-\mu _{-})}{4\xi }\left( -\frac{d^{2}}{dr^{2}}+%
\frac{g}{r^{2}}\right) +\frac{\mu _{-}-\mu _{+}}{4}\left( 1+2r\frac{d}{dr}%
\right) +\frac{(\mu _{0}+\mu _{-}+\mu _{-})}{4}\xi r^{2},
\end{equation}%
which under the action of 
\begin{equation}
\eta =\exp {\left[ \frac{1}{\sqrt{1-4\lambda ^{2}}}\mathrm{ArcTanh}\left( 
\frac{(\mu _{-}-\mu _{+})\sqrt{1-4\lambda ^{2}}}{\mu _{-}+\mu _{+}-2\lambda
\mu _{0}}\right) [K_{0}+\lambda (K_{+}+K_{-})]\right] }
\end{equation}%
transforms into 
\begin{equation}
h=\frac{a+b}{\xi }\left( \frac{d^{2}}{dr^{2}}-\frac{g}{r^{2}}\right) +\left( 
\frac{1+2\lambda }{1-2\lambda }a+3b\right) \xi r^{2}
\end{equation}%
with parameters given by 
\begin{eqnarray}
a &=&\frac{1}{2(1+2\lambda )}\sqrt{\mu _{+}\mu _{-}-\lambda \mu _{0}(\mu
_{+}+\mu _{-})+\lambda ^{2}(\mu _{0}^{2}+(\mu _{-}-\mu _{+})^{2})}, \\
b &=&\frac{\mu _{0}-2\lambda (\mu _{+}+\mu _{-})}{4(1-4\lambda ^{2})}.
\end{eqnarray}

As already mentioned mild varations of the representation (\ref{q2}), (\ref%
{q3}) can be used to obtain multi-particle systems, such as Calogero models.
An easier multi-particle model is the two-mode Bose-Hubbard model \cite{Eva}
or the description of a charged particle in a magnetic field \cite{Nova},
which result when taking as representation 
\begin{equation}
K_{0}=\frac{1}{2}\left( a_{2}^{\dag }a_{2}-a_{1}^{\dag }a_{1}\right) ,\qquad
\qquad K_{+}=a_{2}^{\dag }a_{1},\qquad \qquad K_{-}=a_{1}^{\dag }a_{2},
\label{k12}
\end{equation}%
where the $a_{i}^{\dag },a_{i}$ are the creation and annihilation of the $i$%
-th bosonic particle. It is straightforward to apply the above programme
also to this type of system.

As a variation of the above idea we may also study multi-particle $\mathcal{%
PT}$\textit{-}symmetric Hamiltonians, for which we do not mix different
particle types implicitly within $su(1,1)$-generators, i.e. taking direct
sums of Fock speces, but consider instead systems of the type $su(1,1)\oplus
su(1,1)$, such as 
\begin{eqnarray}
H_{m} &=&\mu _{0}^{(1)}K_{0}^{(1)}+\mu _{+}^{(1)}K_{+}^{(1)}+\mu
_{-}^{(1)}K_{-}^{(1)}+\mu _{0}^{(2)}K_{0}^{(2)}+\mu
_{+}^{(2)}K_{+}^{(2)}+\mu _{-}^{(2)}K_{-}^{(2)}+\mu
_{00}K_{0}^{(1)}K_{0}^{(2)}  \notag \\
&&+\mu _{+-}K_{+}^{(1)}K_{-}^{(2)}+\mu _{-+}K_{-}^{(1)}K_{+}^{(2)}+\mu
_{++}K_{+}^{(1)}K_{+}^{(2)}+\mu _{--}K_{-}^{(1)}K_{-}^{(2)}++\mu
_{+0}K_{+}^{(1)}K_{0}^{(2)}  \notag \\
&&+\mu _{0-}K_{0}^{(1)}K_{-}^{(2)}+\mu _{0+}K_{0}^{(1)}K_{+}^{(2)}+\mu
_{-0}K_{-}^{(1)}K_{0}^{(2)}
\end{eqnarray}%
with the superscripts in the $K^{(i)}$ indicate the particle type. We may
start with an ansatz of a similar type 
\begin{equation}
\eta =\exp [2\varepsilon _{1}(K_{0}^{(1)}+\lambda _{1}K_{+}^{(1)}+\lambda
_{1}K_{-}^{(1)})+2\varepsilon _{2}(K_{0}^{(2)}+\lambda
_{2}K_{+}^{(2)}+\lambda _{2}K_{-}^{(2)})]
\end{equation}%
and it is then straightforward to show that the constraints 
\begin{equation}
\mu _{00}=\frac{\mu _{++}}{\lambda _{1}\lambda _{2}},\qquad \mu _{-+}=\mu
_{+-}=\mu _{--}=\mu _{++}
\end{equation}%
\begin{equation}
\mu _{0+}=\mu _{0-}=\frac{\mu _{++}}{\lambda _{1}},\qquad \mu _{+0}=\mu
_{-0}=\frac{\mu _{++}}{\lambda _{2}},
\end{equation}%
with 
\begin{equation}
\frac{\tanh 2\theta _{i}}{\theta _{i}/\varepsilon _{i}}=\frac{\mu
_{-}^{(i)}-\mu _{+}^{(i)}}{\mu _{-}^{(i)}+\mu _{+}^{(i)}-2\lambda _{i}\mu
_{0}^{(i)}}\qquad \text{for }i=1,2
\end{equation}%
convert the Hamiltonian $H_{m}$ into a Hermitian one. Note that despite the
fact that in $H_{m}$ we have an interaction between different particle types
the constraints are identical to the ones in the linear case for individual
\ particles. In fact, $H_{m}$ is indeed linear in $K^{(1)}$ and $K^{(2)}$
and the terms involving the products of $K^{(1)}$ and $K^{(2)}$ operators
are Hermitian. Adding some genuinely bilinear combinations in the
Hamiltonian is expected to generate a more intricate structure sheding light
on interacting spins etc, but we leave this for future investigations as it
goes beyond the simple comment we intended to make in this section regarding
explicit realisations.

There are of course many more realisations we could present. We finish by
mentioning the famous BCS-Hamiltonian \cite{BCS}, which can be expressed in
terms of many copies of algebra (\ref{comK}) and plays a key role in the
theory of superconductivity%
\begin{equation}
H_{\text{BCS}}=\sum\limits_{n,\sigma }\varepsilon _{n}a_{n\sigma }^{\dagger
}a_{n\sigma }-\sum\limits_{n,n^{\prime }}I_{n,n^{\prime }}a_{n\downarrow
}^{\dagger }a_{n\uparrow }^{\dagger }a_{n^{\prime }\downarrow }a_{n^{\prime
}\uparrow }.
\end{equation}%
Here $a_{n\sigma }^{\dagger }$ and $a_{n\sigma }$ are creation and
annihilation operators of electrons with spin $\sigma $ in a state $n$,
respectively. The $\varepsilon _{n}$ are eigenvalues of the one-body
Hamiltonian and the $I_{n,n^{\prime }}$ are matrix elements of short range
electron-electron interaction.

\section{Conclusions}

The central aim of this manuscript was to analyse systematically $\mathcal{PT%
}$-symmetric Hamiltonians of Lie algebraic type with regard to their
quasi-Hermitian solvability properties. We have considered Hamiltonians of $%
sl_{2}(\mathbb{R})$-Lie algebraic type and for some specific cases we
constructed a similarity transformation together with isospectral Hermitian
counterparts. We indicated the difficulty these types of Hamiltonians pose
with regard to the outlined programme, mainly due to the feature that the
Hermitian conjugation does not close within the set of $sl_{2}(\mathbb{R})$%
-generators. Nonetheless, for specific realisations of the algebra the
outlined programme may be carried out explicitly.

Considering Hamiltonians of $su(1,1)$-Lie algebraic type instead circumvents
these issues and we were able to construct systematically exact solutions
for metric operators, which are of exponential form with arguments linear in
the $su(1,1)$-generators. Our solutions fall into various subcases and are
characterized by the constraints on the coupling constants in the model. In
several cases we used the square root of the metric operator to construct
the corresponding similarity transformation and its Hermitian counterparts.
Alternatively we constructed the energy spectrum together with their
corresponding eigenfunctions by means of generalized Bogoliubov
transformations, which map the original Hamiltonians onto harmonic
oscillator type Hamiltonians. The comparison between these two approaches
exhibits agreement in some cases, but the overlap is not complete and we can
obtain SQH-models which can not be mapped to a harmonic oscillator type
Hamiltonian by means of generalized Bogoliubov transformations and vice
versa. On one hand this is probably due to our restrictive ansatz for the
operator $\eta $ by demanding it to be Hermitian and in addition assuming it
to be of exponential form with arguments linear in the $su(1,1)$-generators.
On the other hand we could of course also make a more general ansatz for the
"target Hamiltonian"\ in the generalized Bogoliubov transformation approach.

There are some obvious omissions and further problems resulting from our
analysis. It would be desirable to complete the programme for more general
operators $\eta $ and different types of Bogoliubov transformations for the
Hamiltonians of rank one Lie algebraic type. With regard to some concrete
physical models, e.g.~the Regge model or other types mentioned in \cite{PEGA}%
, it is necessary to investigate Hamiltonians with a mixture between $%
su(1,1) $- and $sl_{2}(\mathbb{R})$-generators. Naturally systems related to
higher rank algebras constitute an interesting generalisation.

\noindent \textbf{Acknowledgments}: P.E.G.A. is supported by a City
University London research studentship.


\end{document}